\documentclass[aps,twocolumn,showpacs,superscriptaddress,amsmath]{revtex4}
\usepackage{graphicx}
\newcommand{\bea}{\begin{eqnarray}}
\newcommand{\beq}{\begin{equation}}
\newcommand{\eea}{\end{eqnarray}}
\newcommand{\eeq}{\end{equation}}
\begin{document}
\title{Dynamical screening of the Coulomb interaction for two confined electrons
in a magnetic field}
\author{ N. S. Simonovi\'c}
\affiliation{Institute of Physics, P.O. Box 57, 11001 Belgrade,
Serbia}
\author{R. G. Nazmitdinov}
\affiliation{Departament de F{\'\i}sica, Universitat de les Illes Balears, E-07122
Palma de Mallorca, Spain}
\affiliation{Bogoliubov Laboratory of Theoretical Physics,
Joint Institute for Nuclear Research, 141980 Dubna, Russia}
\date{\today}
\begin{abstract}
We show that a difference in  time scales
of vertical and lateral dynamics permits one  to analyze
the problem of interacting electrons confined in an axially symmetric
three-dimensional potential with a lateral oscillator confinement
by means of
the effective two-dimensional Hamiltonian with a screened
Coulomb interaction. Using an adiabatic approximation based on
action-angle variables, we present solutions for the effective
charge of the Coulomb interaction (screening) for
a vertical confinement potential simulated by parabolic, square, and
triangular wells. While for the parabolic potential the solution
for the effective charge is given in a closed anlytical form,
for the other cases similar solutions can be easily calculated numerically.
\end{abstract}
\maketitle

\section{Introduction}

It is well known that there is a restricted class of exactly solvable
problems in quantum mechanics. Such examples serve as  paradigms to illustrate
fundamental principles or/and new methods in their respective fields.
This is especially important for finite systems, where approximative methods are
indispensable to a many-body problem.  In particular, two-electron systems play
an important role in understanding of electron correlation
effects because their eigenstates can be obtained very accurately, or in some cases,
exactly.

The most popular model to study the electronic exchange-correlation energy
in a density functional theory is the Hookean two-electron atom (HA). The basic
HA is two electrons interacting by the Coulomb potential  but bound to a nucleus
by a harmonic potential that mimics a nuclear-electron attraction.
For certain values of the
confinement strength, there exist exact solutions for the HA ground state
\cite{kai,taut}. When the HA is placed in a perpendicular magnetic field,
Taut provided analytical solutions for a two-dimensional (2D) HA at particular values
of the magnetic field \cite{taut2}. This model can be equally viewed as a model
2D quantum dot (QD). Recent progress in semiconductor technology made it possible
to fabricate and probe such confined system at different values of the magnetic
field \cite{as,kou}. Consequently, it has stimulated numerous theoretical
studies on two-electron QDs, so-called "artificial atoms" (see for a recent
review \cite{RM}). Being a simple nontrivial system, QD He poses a significant
challenge to theorists. For example,
using a 2D He QD model, one is able to reproduce a general trend for
the first singlet-triplet (ST) transitions observed in two-electron QDs
under a perpendicular magnetic field.
However, the experimental positions of the ST transition
points are systematically higher \cite{kou,ni}.
The ignorance of the third dimension is the most evident source
of the disagreement, especially, in vertical QDs \cite{ron,bruce,nen3}.

Although accurate numerical results for QD He can be obtained
readily, analytical results are still sought, because
they provide the physical insight into numerical calculations.
Moreover, analytical results could
establish a theoretical framework for accurate analysis of
confined many-electron systems, where the exact treatment of the three-dimensional (3D) 
case becomes computationally intractable. The purpose  of the present paper is to introduce
a consistent approach which enables one to reduce the  3D
Coulomb problem to the effective 2D one, without loosing major effects related
to the QD's thickness.
In Section II we develop the concept of the effective  charge
of the Coulomb potential based on the adiabatic approximation \cite{RRM} for axially
symmetric 3D systems.
The calculations of the effective charge
for a Hamiltonian with a vertical confinement approximated by
parabolic, square, and triangular well potentials are presented in Section III.
Finally, we will show in Section IV that recent experimental data \cite{ni}
can be successfully reproduced within a
2D approximation but with additionally screened Coulomb interaction
due to the thickness of the sample.
The main outcomes are summarized in Section V. Three Appendixes provide
some technical details of the calculations.

\section{Basic remarks}
\subsection{Model}
\label{mod}
The system Hamiltonian for the case of a magnetic field B along a symmetry axis z
reads
\beq
\label{ham}
H = \sum_{i=1}^2 \bigg[ \frac{1}{2m^*\!}\,
\Big({\bf p}_i - \frac{e}{c} {\mathbf A}_i \Big)^{\! 2}
+ U({\mathbf r}_i) \bigg]
+ V_C+H_{\it spin}.
\eeq
Here the term $V_C=k/{\vert{\mathbf r}_1 \!- {\mathbf r}_2\vert}$ with
$k = e^2/4\pi\epsilon_0\epsilon_r$ describes the
Coulomb repulsion between electrons and
$H_{\it spin}=g^*\mu_B({\bf s}_1+{\bf s}_2)¿{\bf B}$ is the Zeeman term,
where $\mu_B=|e|\hbar/2m_ec$ is the Bohr magneton.
 For the magnetic field we choose the vector potential with
gauge ${\mathbf A}_i = \frac{1}{2} {\mathbf B} \times
{\mathbf r}_i = \frac{1}{2}B(-y_i,x_i,0)$. The confining potential is
approximated by a 2D circular harmonic oscillator (HO) in $xy$-plane 
and the vertical confinement
$V_z$: $U({\mathbf r}_i) = m^*
\omega_0^2\,\rho_i^2/2 + V_z(z_i)$; $r_i^2 = \rho_i^2+z_i^2$,
$\rho_i^2 = x_i^2 \!+ y_i^2$ and
$\hbar\omega_0$ is the energy scale of confinement in the $xy$-plane.
Below we analyze different forms for the vertical confinement $V_z$.

For our analysis it is convenient to use
cylindrical coordinates $(\rho,\phi,z)$.
Also, we separate the Hamiltonian (\ref{ham})
on several terms:
$H = H_{0}+H_{z}+V_C+H_{\it spin}$, where the term
$H_0=\sum_{i=1}^2h_i $ consists of the contributions related only to the lateral
dynamics ($xy-$plane) of non-interacting electrons
\beq h_i = t_i + v_i - \omega_L
l_{z_i}=\frac{p_{\rho_i}^2}{2m^*} +
\biggl(\frac{l_{z_i}^2}{2m^*\rho_i^2} +
\frac{1}{2}\,m^*\Omega^2\rho_i^2\biggr) -\omega_L l_{z_i}\;.
\label{tv}
\eeq
Here, the effective lateral confinement frequency
$\Omega = (\omega_0^2 + \omega_L^2)^{1/2}$
depends on the magnetic field by means of the Larmor frequency
$\omega_L=|e|B/2m^*c$; $l_{z_i} \equiv p_{\phi_i}$ is the
$z$-component of the angular momentum of the $i$-th electron.
The eigenstates of the single-particle Hamiltonian (\ref{tv})
are well-known Fock-Darwin states (cf \cite{qd}).
The motion of noninteracting electrons in the $z$-direction is described
by the Hamiltonian
$H_z=\sum_i(p_{z_i}^2/{2m^*}+ V_z(z_i))$.
Since the magnetic field is directed along the z axis,
the Zeeman term is $H_{\it spin}=g^*\mu_B S_zB$.
This term is not important for our analytical study  and will be taken
into account only in numerical analysis of experimental data.

\subsection{Adiabatic approximation}
\label{adapp}

For typical QDs ($\hbar\omega_0 \sim 3$\,meV) the contribution of
the Coulomb interaction to the total energy is comparable to the
confinement energy at zero magnetic field \cite{kou}. Evidently, the
standard perturbation theory is not valid in this case.
In real samples  the confining potential in the
$z$-direction is much stronger than in the $xy$-plane. It results in
different time scales (see below) and this allows one to use the adiabatic
approach \cite{RRM}.
To lowest order the adiabatic approach consists of averaging
the full 3D Hamiltonian over the angle variables
$\theta_{z_i} = \omega_{z_i} t$ (fast variables) of the
unperturbed motion $(k = 0)$ after rewriting the
$(z_i, p_{z_i})$ variables in terms of the action-angle variables
$(J_{z_i}, \theta_{z_i})$.
As a result, the motion effectively
decouples into an unperturbed motion in the vertical direction
governed by the potential $\sum_i V(J_{z_i}, \theta_{z_i})$
and into the lateral motion governed by the effective potential
\beq
V_\mathrm{eff}(\{x,y\};\{J_{z}\}) = \sum_i v_i
+ V_\mathrm{int}^\mathrm{eff}(\rho;J_{z_1},J_{z_2}),
\eeq
where $v$ is defined in Eq.(\ref{tv}), $\rho = [(x_1-x_2)^2+(y_1-y_2)^2]^{1/2}$,
and
\bea
&&V_\mathrm{int}^\mathrm{eff}(\rho;J_{z_1},J_{z_2}) =
\label{vceff}\\
&&\frac{1}{(2\pi)^2}\int_{0}^{2\pi}\!\!d\theta_{z_1}
\int_{0}^{2\pi}\!\!d\theta_{z_2}
V_C(\rho,z_1(J_{z_1},\theta_{z_1})\!-\!z_2(J_{z_2},
\theta_{z_2}))
\nonumber
\eea
is the effective electron-electron interaction that contains the memory on
$z$ dynamics through integrals of motion $J_{z_i}$. The effective
interaction affects, therefore, only the dynamics in the lateral plane,
where the confining potential is the parabolic one (see Eq.(\ref{tv})).
Hence, the effective Hamiltonian
for two-electron QD reads as
\beq
H_\mathrm{eff} = H_0
+ E_{z} + V_\mathrm{int}^\mathrm{eff},
\label{effham}
\eeq
where $E_z=\sum_i\varepsilon_i$ and $\varepsilon_i$
is the electron energy of the unperturbed motion
in the vertical direction.

Our {\it ansatz} consists in the consideration of
the effective interaction (\ref{vceff}) in the form
$V_\mathrm{int}^\mathrm{eff} = k f(\rho)/\rho$. Then one can define the
effective 2D Coulomb interaction
\beq
V_C^\mathrm{eff}=
\frac{k_\mathrm{eff}}{\rho},
\eeq
where the effective charge
is the mean value of the factor $f(\rho)$ upon the nonperturbed lateral
wave functions, i.e.,
\beq
k_\mathrm{eff} = k\langle f(\rho)\rangle \equiv \langle
\rho\,V_\mathrm{int} ^\mathrm{eff}(\rho)\rangle.
\label{effch}
\eeq
In contrast to a standard 2D consideration of the bare Coulomb potential
($V_C=k/\rho$) in QDs \cite{kou,RM}, in our approach the electron dynamics is
governed by the same potential but with the additional effective (screened)
charge due to the QD's thickness.
One of the main advantages of this approach is that the
interaction matrix elements can be expressed in an analytical form.
Thus, we shall use (diagonalize) the effective
Hamiltonian (\ref{effham}) with the effective Coulomb interaction
$V_C^\mathrm{eff}$, i.e.,
$V_\mathrm{int}^\mathrm{eff}\Rightarrow V_C^\mathrm{eff}$.

According to the Kohn theorem \cite{kohn} the center of mass (CM) and
the relative motion of the 2D system described by the Hamiltonian
(\ref{effham}) are separated and the mean value in
Eq.(\ref{effch}) can be evaluated using the Fock-Darwin states for
the relative motion
\begin{equation}
\psi_{n_\rho m}(\rho,\varphi)=
\frac{e^{im\varphi}}{\sqrt{2\pi}} R_{n_\rho m}(\rho).
\label{fd}
\end{equation}

This state is eigenfunction of the operator $l_z$ with eigenvalue
$m$ and the radius-dependent function with a radial quantum number
$n_\rho$ has the form
\begin{equation}
R_{n_\rho m}(\rho) =
\sqrt{\frac{2\mu\Omega n_\rho !}{\hbar(n_\rho+|m|)!}}\,
\xi^{|m|} e^{-\xi^2/2} L_{n_\rho}^{|m|}(\xi^2),
\label{fdl}
\end{equation}
where $\mu = m^*/2$ is the reduced mass, $\xi =
(\mu\Omega/\hbar)^{1/2}\rho$ and $L_{n_\rho}^{|m|}$ denotes the
Laguerre polynomials \cite{LL}. For the lowest states (with
different values of the quantum number $m$ but with the radial
quantum number $n_\rho = 0$) one obtains for the effective charge
\beq
k_\mathrm{eff} = \frac{2}{|m|!}
\biggl(\frac{\mu\Omega}{\hbar}\biggr)^{|m|+1} \int_0^\infty
e^{-\mu\Omega\rho^2/\hbar} \rho^{2|m|+2}\,
V_\mathrm{int}^\mathrm{eff}(\rho)\,d\rho\;.
\label{intk}
\eeq

\section{Effective charge}
\label{efc}

\subsection{Parabolic potential}
\label{ecpar}

\begin{figure}
\includegraphics[width=0.44\textwidth]{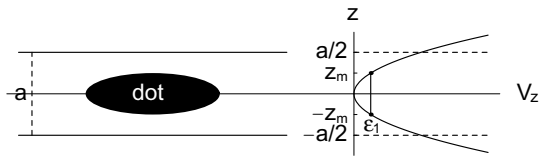}
\caption{
Left: the localization of QD in the layer of the thickness $\mathrm{a}$.
Right: the schematic representation of the position of zero-point
motion in the parabolic confinement relative to the layer thickness.
}
\label{fig1}
\end{figure}

To account for effect of localization of
the dot in the layer of thickness $\mathrm{a}$,
let us first consider  a 3D model with a
vertical confinement  approximated by a parabolic potential
$V_z(z_i) = \frac{1}{2}\,m^*\omega_z^2\,z_i^2$ (see Fig.~\ref{fig1}).
Due to  the Kohn theorem \cite{kohn} the CM and the
relative motions are separated.
The solution for the CM motion is well known (cf \cite{qd}).
It is not important for our discussion, since the CM dynamics does not affect
the electron interaction.
The 3D Hamiltonian for the relative
motion of two electrons has the form
\beq
 H_\mathrm{rel}=h_\mathrm{rel}+\frac{k}{r}+\frac{p_z^2}{2\mu}+
 \frac{\mu\omega_z^2 z^2}{2}.
\label{relham}
\eeq
The term $h_\mathrm{rel}$ is defined by Eq.(\ref{tv})
in which the effective electron mass is replaced
by the reduced mass $\mu$;
all single-electron variables (with index $i$) are
replaced by the corresponding variables for relative motion
(without indices).

After rewriting the $(z, p_z)$ variables in terms of the
action-angle variables $(J_z, \theta_z)$
\beq
z  = \sqrt{\frac{2J_z}{\mu\omega_z}}\,\sin\theta_z,
\quad p_z = \mu\dot{z},
\label{ztz}
\eeq
we integrate out of the fast variable, i.e.,
average the Hamiltonian (\ref{relham}) over
the angle $\theta_z$. As a result,
the effective interaction potential (see also Appendix \ref{app1}) is
\beq
V_\mathrm{int}^\mathrm{eff}(\rho;J_z) = \frac{2k}{\pi\rho}\,K\biggl(-\frac{2
J_z}{\mu \omega_z \rho^2}\biggr),
\label{effvcpar}
\eeq
where $K(x)=\int_0^{\pi/2}(1-x\sin^2\theta)^{-1/2} d\theta$
is the complete elliptic integral of the first kind (see
\cite{abr}). This integral is well defined for all values of
$\rho$.

The effective Hamiltonian for the relative motion is
\bea
&H_\mathrm{rel}^\mathrm{eff} = h_\mathrm{rel}+
V_\mathrm{int}^\mathrm{eff}(\rho;J_z)+ E_z^\mathrm{rel}
\label{scaleffen}\\
&E_z^\mathrm{rel} =\omega_z J_z= \hbar\omega_z(n_z + 1/2).
\label{ez}
\eea
To replace the effective electron-electron potential
$V_\mathrm{int}^\mathrm{eff}$ by the Coulomb-type
$V_C^\mathrm{eff}$ in Eq.(\ref{scaleffen}),
we must determine the effective charge.
Taking into account the definitions Eqs.(\ref{effch}), (\ref{ez})
and the result Eq.(\ref{effvcpar}), one obtains the following for
the effective charge
\beq
k_\mathrm{eff} = \frac{2k}{\pi} \biggl\langle n_\rho,m\bigg|
K\biggl(-\frac{\hbar\,(2n_z + 1)}{\mu \omega_z
\rho^2}\biggr)\biggr|n_\rho,m\biggr\rangle,
\label{k1}
\eeq
where $|n_\rho,m\rangle$ are the Fock-Darwin states for the
relative motion, Eqs.~(\ref{fd}),(\ref{fdl}).

For the lowest states ($n_\rho = n_z = 0$) one
reduces Eq.(\ref{k1}) to the form of integral (\ref{intk}).
As a result, the effective charge can be expressed in terms of the Meijer
G-function \cite{grad}
\begin{equation}
k_\mathrm{eff} = \frac{k}{\pi |m|!}\,
      G^{2,2}_{2,3}
  \biggl(
\frac{\Omega}{\omega_z} \biggr|
 \begin{array}{c}
{1/2}\ \ {1/2}\\
0\,\,{m\!+\!1}\,\,0\\
\end{array} \biggr).
 \label{kaap}
\end{equation}

Guided by the adiabatic approach, it is
instructive to compute the effective charge by dint of quantum-mechanical
mean value of the Coulomb term in the 3D oscillator state
$|n_\rho,m\rangle |n_z\rangle$
\beq
k_{\mathrm{eff}} = \langle \langle \rho\,V_C(\rho,z)\rangle \rangle =
k\,\langle \langle (1+z^2/\rho^2)^{-1/2}\rangle \rangle \,.
\label{kqm}
\eeq
Here, $|n_z\rangle$ is a normalized one-dimensional harmonic oscillator
wave function \cite{LL}. Since the lateral extension exceeds the
thickness of the QDs by several times, one may suggest to consider
the ratio $(z/\rho)^2$ as a  small parameter of theory. Note, however, that
the averaging over the 3D oscillator
state $|n_\rho,m\rangle |n_z\rangle$ implies the application of the first order
perturbation theory for calculation of the contribution of the Coulomb interaction
in QDs. For $n_\rho = n_z = 0$ one obtains
\begin{widetext}
\beq
k_\mathrm{eff} =
k\frac{2}{|m|!} \biggl(\frac{\mu\Omega}{\hbar}\biggr)^{|m|+1}
\sqrt{\frac{\mu\omega_z}{\pi\hbar}} \int_0^\infty
K_0\biggl(\frac{\mu\omega_z\rho^2}{2\hbar}\biggr)\,
e^{\mu(\frac{1}{2}\omega_z-\Omega)\rho^2/\hbar}
\rho^{2|m|+2}\,d\rho,
\label{k1a}
\eeq
\end{widetext}
where $K_0$ is the modified Bessel function of the 2nd kind.
One observes that in both definitions of the effective charge
Eqs.(\ref{k1}),(\ref{k1a}) there is a contribution of the
electron dynamics along  the coordinate $z$. Below we will compare
true contagion of both definitions upon the interpretation
of the experimental data.

For small and relatively large values of the magnetic field
$0 < \Omega/ \omega_z< 1$, available in
experiment, the solution for the last integral  can be expressed
in terms of the hypergeometric functions

\begin{widetext}
\bea
k_\mathrm{eff} &=& k\frac{2^{2|m|+1}}{\sqrt{\pi}|m|!}
\bigg(\frac{\Omega}{{\omega}_z} \bigg)^{|m|+1} \Bigg[
\Gamma\bigg(\frac{2|m|\!+\!3}{4}\bigg)^2
{_2F_1}\bigg(\frac{2|m|\!+\!3}{4},\frac{2|m|\!+\!3}{4},\frac{1}{2},
\Big(1-2\frac{\Omega}{{\omega}_z}\Big)^2\bigg) +\nonumber
\\
&&2\bigg(1-2\frac{\Omega}{{\omega}_z}\bigg)
\Gamma\bigg(\frac{2|m|\!+\!5}{4}\bigg)^2
{_2F_1}\bigg(\frac{2|m|\!+\!5}{4},\frac{2|m|\!+\!5}{4},\frac{3}{2},
\Big(1-2\frac{\Omega}{{\omega}_z}\Big)^2\bigg) \Bigg].
\label{k2a}
\eea
\end{widetext}

\subsection{Infinite square well}
\label{sqw}

A hard wall potential is among popular models for the confinement in QDs.
In case of the vertical confinement, $V_z$ is
the one-dimensional infinite square well
\beq V_z = \bigg\{
\begin{matrix}
0, & |z| < a/2 \\
\infty, & |z| \ge a/2
 \end{matrix}\quad,
\label{sqwell}
 \eeq
where $a$ is the thickness of the layer in which the
dot is created (see Fig.~\ref{fig1}).

Due to uniform motion inside the square well, the angle variable
$\theta_{z_i}$ is proportional to the coordinates
$z_i$ (see Eq.~(\ref{awell}) in Appendix \ref{appA})
and, therefore,
\bea
&&V_\mathrm{int}^\mathrm{eff}= \frac{1}{\pi^2}\int_{-\pi/2}^{\pi/2}
d\theta_{z_1} \int_{-\pi/2}^{\pi/2} d\theta_{z_2} V_C(\rho,z)
\nonumber\\
&=& \frac{k}{a^2}\int_{-a/2}^{a/2} dz_1 \int_{-a/2}^{a/2} dz_2
\,[\rho^2 + (z_1 - z_2)^2]^{-1/2}.
\eea
Integrating over the coordinates $z_1$ and $z_2$ between the
walls of the potential (\ref{sqwell})
one obtains the effective Coulomb interaction
$V_\mathrm{int}^\mathrm{eff}= k{\cal A}(\xi)/a$, where $\xi = \rho/a$ and
\beq
{\cal A}(\xi) =
\biggl[2\xi - 2\sqrt{1+\xi^2} +
\ln\biggl(\frac{\sqrt{1+\xi^2}+1}{\sqrt{1+\xi^2}-1}\biggr)\biggr]\;.
\eeq
The effective Hamiltonian now reads
\beq
 H_\mathrm{eff} = \sum_{i=1}^2 \biggl[h_i+
 \frac{\pi^2\hbar^2 n_{z_i}^2}{2m^*a^2}\biggr] + V_\mathrm{int}^\mathrm{eff},
\eeq
where the term $h_i$ is determined by Eq.(\ref{tv}) and we use
Eq.(\ref{ebox}) for the contribution of the vertical
confinement. In contrast to the parabolic potential,
here $V_\mathrm{int}^\mathrm{eff}$ does not depend on the quantum numbers $n_{z_i}$.
In order to replace
$V_\mathrm{int}^\mathrm{eff} \to V_C^\mathrm{eff}$ we must
define the effective charge. Similar to the previous case,
the effective charge is the mean value $\langle
\rho\,V_\mathrm{int}^\mathrm{eff}(\rho)\rangle$ in the Fock-Darwin states.
For the lowest states with different $m$ we obtain
\beq
k_\mathrm{eff} =
k\frac{2 b^{|m|+1}}{|m|!} \int_0^\infty
e^{-b\,\xi^2} \xi^{2|m|+2}{\cal A}(\xi)\,d\xi \; ,
\label{k2}
\eeq
where $b = \mu\Omega a^2/\hbar$.

\subsection{Triangular well}
\label{tri}

Let us consider the well potential characterized by an infinitely high barrier
for $z < 0$ and a linear potential $V_z(z)=eFz$ for $z > 0$;
 the product of the electron charge $e$ and an electric field
$F$ is assumed to be positive:
\beq V_z(z) = \bigg\{
\begin{matrix}
\infty, & z < 0 \\
eFz, & z \ge 0
\end{matrix}\;.
\label{triwell}
 \eeq
This potential is a simple realistic
description of the potential well at doped heterojunction or/and
in the case when an external voltage is applied to the (top and
bottom) electrodes of the sample in which the dot is created.

By means of the relation Eq.(\ref{zang}) (see Appendix \ref{appB})
between the coordinates $z_i$ and the angle variables $\theta_{z_i}$,
for the case when both electrons occupy the same energy level
$\varepsilon$, the effective Coulomb term is
\bea
&&V_\mathrm{int}^\mathrm{eff} = \frac{1}{\pi^2}\int_{0}^{\pi}
d\theta_{z_1} \int_{0}^{\pi} d\theta_{z_2} V_C(\rho,z)
\label{vetr}
\\
&&= k\int_0^1 d\tilde\theta_1 \int_0^1 d\tilde\theta_2
\bigl[\rho^2 + z_m^2\,(\tilde\theta_1 -
\tilde\theta_2)^2(2-\tilde\theta_1 - \tilde\theta_2)^2
\bigr]^{-1/2} \nonumber\\
&&=\frac{k}{z_{m}}\int_{-1}^0 d\xi_1 \int_{-\xi_1-2}^{\xi_1} d\xi_2
\bigl(\xi_3^2 + \,\xi_1^2\,\xi_2^2\bigr)^{-1/2}=
\frac{k}{z_{m}}{\cal B}(\xi_3),
\nonumber
\eea
where $\xi_1 = \tilde\theta_1 - \tilde\theta_2$, $\xi_2 =
\tilde\theta_1 + \tilde\theta_2 - 2$, $\xi_3=\rho/z_m$,
$z_m=\varepsilon/eF$ and $\tilde\theta_i \equiv \theta_{z_i}/\pi$.

According to Eq.(\ref{etr}), the lowest $(n_z=1)$
energy level in the triangular potential is $\varepsilon_1 =
c_1(\hbar eF/\sqrt{2m^*})^{2/3}$, where
$c_1 = (9\pi/8)^{2/3}\approx 2.32$.
 The parameters can be selected to ensure the corresponding wave function
(the Airy function, the first zero of this function occurs at $c_1=2.338$;
see, for example, Ref.\onlinecite{LL}) will be
extended until the desired length in the vertical direction.
Thus, in the adiabatic approximation the effective Hamiltonian has the form
\beq H_\mathrm{eff} = H_0 + 2\varepsilon_1 +
V_\mathrm{int}^\mathrm{eff}(\rho) \;,
\eeq
where the term $H_0$ is defined in subsection \ref{mod}.
As above, we replace $V_\mathrm{int}^\mathrm{eff} \to
V_C^\mathrm{eff}$ and calculate the effective charge.
For the states with $n_\rho = 0$
(the Fock-Darwin states for the relative motion) we obtain the
effective charge (see Eq.(\ref{effch}))
\beq
k_\mathrm{eff} = k\frac{2 b^{|m|+1}}{|m|!} \int_0^\infty
d\xi_3\, e^{-b\,\xi_3^2} \xi_3^{2|m|+2}{\cal B}(\xi_3)\;,
\label{ktr}
\eeq
where $b=\mu\Omega{z_m}^2/\hbar$ and $z_m = \varepsilon_1/eF$.
We recall that the parameter $z_m$ is the distance between turning points.
It is similar to the parameter $a$ in the square well potential.
The integral over the variable $\xi_2$ in the expression for ${\cal B}$
(Eq.(\ref{vetr}))
can be evaluated analytically but the
other two  in Eqs.(\ref{vetr}) and (\ref{ktr}) must be calculated numerically.

\section{Discussion of results}

\begin{figure*}[ht]
\includegraphics[height=0.28\textheight,clip]{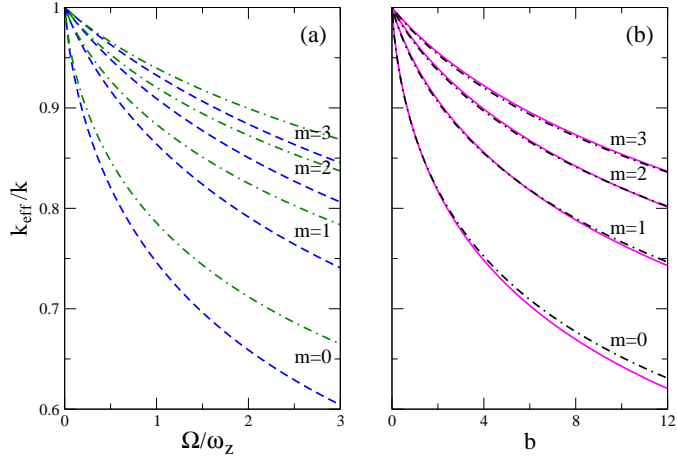}
\caption{(Color online)
The ratio $k_\mathrm{eff}/k$ for the lowest states with
different $m$ for the vertical confinement approximated by:
(a) the parabolic potential; (b) the hard
wall (square/triangular well) potentials.
(a) The results obtained in the adiabatic approximation
and by means of the plain quantum-mechanical averaging procedure
(see subsection \ref{ecpar}) are
connected by dashed (blue) and dotted-dashed (green) lines, respectively.
(b) The results for the square
and triangular well are connected by solid (pink) and dotted-dashed lines,
respectively.
}
\label{fig2}
\end{figure*}

In Fig.~\ref{fig2} we present results for
the effective charge (ratio $k_\mathrm{eff}/k$),
calculated for the lowest states ($n_\rho=n_z=0$) with
different $m$, with the aid of: a)Eq.(\ref{kaap})
for the parabolic confinement as a function of the ratio
$\Omega/\omega_z$ (see Fig.~\ref{fig2}(a)); b)Eq.(\ref{k2}) and
Eq.(\ref{ktr}) for the square and triangular well potentials,
respectively, as a function of the variable $b$ (see Fig.~\ref{fig2}(b)).
These results evidently demonstrate that
the inclusion of the vertical dynamics reduces
the Coulomb interaction. In all considered
models for the vertical confinement this effect affects strongly
quantum states with small values of the quantum number $m$.
Such states are major participants in the
ground state transitions at small and intermediate values of the
magnetic field. Therefore, the attenuation of the Coulomb interaction
due to the sample thickness may explain
the disagreement between experimental data and predictions upon
the position of the singlet-triplet transitions based on
calculations with the 2D Coulomb potential (see below).
In other words, the 2D calculations (with the charge $k$) overestimate the
electron correlations in two-electron QDs.

\begin{figure}[bt]
\includegraphics[scale=0.3]{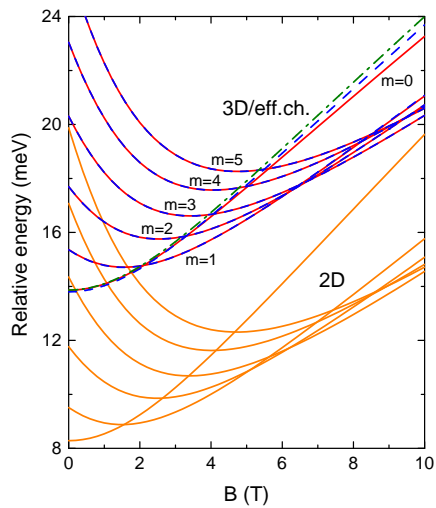}
\caption{Lowest energy levels (only the relative motion is considered)
of the two-electron QD with fully parabolic potential.
The results of the 2D, the 3D and the effective charge (adiabatic)
approximations are connected by solid (orange), solid (red), and
dashed (blue) lines, respectively.
The result of the plain quantum-mechanical averaging for the $m = 0$ level
($k_\mathrm{eff} = \langle \langle \rho V_C\rangle \rangle$) is connected
by the dotted-dashed (green) line.}
\label{fig3}
\end{figure}

Although the results for the effective charge
in the parabolic potential
(adiabatic approach) (see Fig.\ref{fig2}a)
are slightly different in comparison with those of the hard potentials,
a general trend for this charge determined by the magnetic field is similar
(compare Figs.\ref{fig2}a,b).
Notice that the magnitude of the ratio and its evolution
with the increase of the magnetic field
are almost the same in the square and triangular potentials
(see Fig.\ref{fig2}b).
In contrast to the plain quantum-mechanical averaging
procedure (see Eqs.(\ref{kqm}\;-\ref{k2a}))
in the adiabatic approach (see Eq.(\ref{kaap}))
the Coulomb interaction (the ratio $k_\mathrm{eff}/k$)
is reduced stronger (see below).

To elucidate the quality of the adiabatic approach
we use the 3D axially symmetric oscillator model with the
lateral ($\hbar\omega_0 = 3$\,meV) and
the vertical ($\hbar\omega_z = 12$\,meV) confinements and
compare results for energy levels in 2D, 3D and the effective charge
(adiabatic) approximations.
For a numerical analysis of spectral properties we choose the effective
mass $m^*=0.067m_e$, the relative dielectric constant of a semiconductor
$\varepsilon_r=12$ and $|g^*|=0.44$ (bulk GaAs values).
In all three approximations we solve the eigenvalue
problem by means of exact diagonalization of the corresponding Hamiltonian in the
Fock-Darwin basis. We recall that in this basis the Coulomb matrix elements
are given in analytical form (cf \cite{qd}).
One observes a remarkable agreement between results for
the 3D and the effective charge approximations (see Fig.~\ref{fig3}).
In the 2D approximation the Coulomb interaction is much stronger
and, consequently, the evolution of levels with the magnetic field as
well as their absolute values are different from those of the 3D
approximation.

To illuminate the key advantage of the effective charge concept it is
noteworthy to analyze the available
experimental data \cite{ni} within various approaches.
To this aim we will compare the results of calculations of
the additional energy $\Delta \mu=\mu(N)-\mu(N-1)$ \cite{as,kou}
in the 2D approximation (with the effective Coulomb interaction)
and in the 3D approach with a full Coulomb interaction.
Here, the chemical potential of the dot $\mu(N)=E(N)-E(N-1)$
is given by the ground-state energy of the dot $E(N)$ with $N$
and $N-1$ electrons.  To conduct this study,
one needs to estimate the vertical confinement frequency $\hbar\omega_z$
in a real sample. The simplest approach is to use the parabolic model for the
vertical confinement in the layer of thickness $a$ (see Fig.\ref{fig1}).

\begin{figure}[ht]
\includegraphics[height=0.3\textheight,clip]{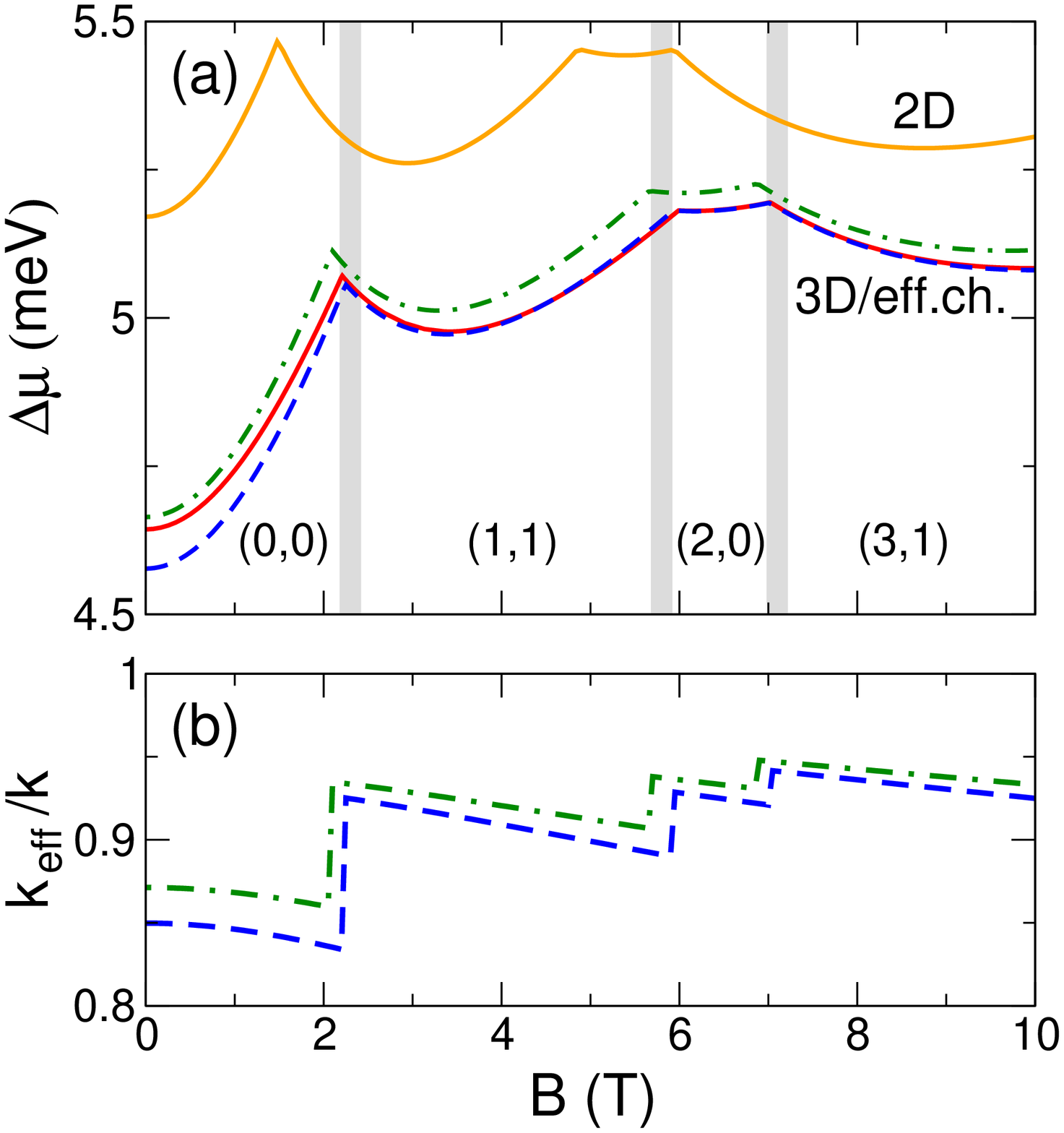}
\caption{(Color online) (a)
The additional energy $\Delta\mu$  as a function
of the magnetic field in the parabolic model  for a two-electron QD. 
The results of calculations
with a lateral confinement only
(the 2D approach, $\hbar\omega_0 = 2.9$\,meV, $|g^*| = 0.3$)
and full 3D approach \cite{nen3} ($\hbar\omega_z = 8$\,meV)
are connected by  solid (orange) and (red) lines, respectively.
The vertical grey lines indicate the position of the experimental
crossings between different ground states \cite{ni}.
Ground states are labeled by
$(m,S)$, where $m$ and $S$ are
the quantum numbers of the operators $l_z$ and the total spin,
respectively.
The results based upon the adiabatic approximation, Eq.(\ref{kaap}), and
the plain quantum-mechanical averaging procedure, Eq.(\ref{k1a}), are connected
by dashed (blue) and dot-dashed (green) lines, respectively.
(b) The ratio $k_\mathrm{eff}/k$  as functions of the magnetic field
based on Eq.(\ref{kaap}) and the plain quantum-mechanical averaging, Eq.(\ref{k1a})
are connected by dashed (blue) and dotted-dashed (green) lines, respectively.}
\label{fig4}
\end{figure}

The thickness of QDs is much smaller in comparison with the
lateral extension. Therefore, the vertical confinement
$\hbar\omega_z$ is much stronger than the lateral confinement
$\hbar\omega_0$ and this fact is, usually, used to justify
a 2D approach for the study of QDs.  However, there is a nonzero contribution from the
vertical dynamics, since the energy level
available for each of the noninteracting electrons in the $z$ direction is
$\varepsilon = \hbar\omega_z(n_z+1/2)$. For the lowest state
$n_z = 0 \Rightarrow \varepsilon_1 = \frac{1}{2}\,
\hbar\omega_z $. By dint of the condition $V_z(\pm z_m)\equiv
m^*\omega_z^2 z_m^2/2 = \varepsilon_1 $ one defines the
turning points: $z_m = \sqrt{\hbar/(m^*\omega_z)}$
(see Fig.\ref{fig1}).
We assume that the distance between turning points should not exceed
the layer thickness, i.e., $2z_m \leq a$. From this inequality
it follows that the lowest limit for the vertical confinement in the layer
of thickness $a$ is
\beq
\hbar \omega_z \geq \frac{4\hbar^2}{m^*a^2}
\label{omz_est}
\eeq
or $b > 2\Omega/\omega_z$. For typical GaAs samples with the
thickness $a$ between 10\,nm and 20\,nm this estimation gives the
minimal value for $\hbar\omega_z$ between 45\,meV and 11\,meV,
respectively. These estimations provide  a genuine cause for
the use of the adiabatic approach in case of QDs, since
$T_z(=2\pi/\omega_z)\ll T_0(=2\pi/\omega_0)$.

Using the "experimental" values for the lateral confinement  and
the confinement frequency $\omega_z$ as a free parameter,
we reproduced successfully with the value $\hbar\omega_z=8$ meV and
$|g^*| = 0.3$  the positions of kinks in the additional energy  for a two-electron QD  
(see Fig.\ref{fig4}a)
\begin{equation}
\Delta \mu = E_\mathrm{rel} - E(1) + E_\mathrm{spin}
\label{dmu}
\end{equation}
in all three samples \cite{ni}  in the 3D axially symmetric oscillator model \cite{nen3}.
Here, $E_\mathrm{rel}=\langle H_\mathrm{rel}\rangle$ is
the relative energy, $E(1)=\hbar\omega_0+\hbar\omega_z/2$ is a single-electron energy
and the Zeeman energy $E_\mathrm{spin} = -|g^*|\mu_B [1-(-1)^m] B/2$ is
zero for the singlet states.
Note that it was found from the Zeeman splitting at high
magnetic field that $|g^*|=0.3$ \cite{ni2}. While at small magnetic field the
results for $|g^*|=0.44$ and $|g^*|=0.3$ are similar, the latter value provides
the best agreement between the full 3D calculations and the experimental data.

We recall that in the 2D approach used by Nishi {\it et al.} \cite{ni}
one encounters the problem of the correct interpretation of
the experimental data (see  Fig.\ref{fig4}a,
a plain lateral confinement $\hbar\omega_0 = 2.9$\,meV).
In the effective charge approximation the
vertical confinement is taken into account with the aid of $k_\mathrm{eff}$ in
the 2D effective Hamiltonian. The remarkable accord between the predictions based on
the results of  Section \ref{ecpar} and the observation confirms the validity
of the suggested concept (see Fig.~\ref{fig4}).
We stress that the effective charge approximation facilitates the calculations
providing the same results as the full 3D calculations.

\begin{figure}[ht]
\includegraphics[height=0.3\textheight,clip]{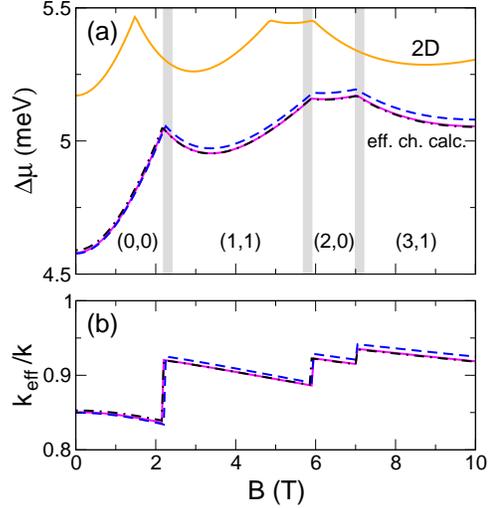}
\caption{(Color online) (a) The magnetic dependence of the
additional energy $\Delta\mu$ for a two-electron QD with the
lateral confinement $\hbar\omega_0 = 2.9$\,meV and $|g^*| = 0.3$.
The results of the pure 2D approach are connected by the solid (orange) line.
The results of the adiabatic approach (effective charge calculations)
are connected: by  the dashed (blue) line for the parabolic model
($\hbar\omega_z = 8$\,meV); by the solid (pink) line for
the square well model ($a = 32.5$\,nm);
by the dotted-dashed (black) line for the triangular well model
($F = 220$\,kV/m). See the text for details.
The vertical grey lines indicate the position of the experimental
crossings between different ground states labeled
by $(m,S)$ \cite{ni}. (b) The ratio
$k_\mathrm{eff}/k$ for three different approximations
of the vertical confinement as a function of the magnetic field.}
\label{fig5}
\end{figure}

Notice that the results based upon the adiabatic approximation are in a
better agreement with the full 3D calculations
in contrast to
those obtained with the aid of the plain quantum-mechanical averaging procedure
for $k_{\mathrm{eff}} = \langle \langle \rho\,V_C(\rho,z)\rangle
\rangle$.  As discussed in Section \ref{adapp},
the adiabatic approach is based
on the effective separation of fast (vertical) and slow (lateral) dynamics
with subsequent averaging procedure. In contrast, the plain quantum-mechanical
averaging  represents a type of perturbation theory based upon the first order
contribution with respect to the ratio $z/\rho$ only. The higher order term may
improve the agreement at small magnetic field, since the vertical dynamics is
non  negligible and affects the lateral dynamics.
The increase of the quantum number $m$, caused by the increase of the
magnetic field strength, reduces the orbital motion of electrons in the
vertical direction. The larger is $m$ the stronger is the centrifugal force,
which induces the electron localization in a plane, and,
therefore, the lesser is importance of the vertical electron
dynamics. In the limit of strong magnetic field (large $m$) the dot becomes more
of a "two-dimensional" system. This explains the improvement of the accuracy
of the plain quantum-mechanical averaging procedure at
large $m$, i.e., for the ground states at high magnetic fields.

The results of different averaging procedures indicate that the accuracy of the
methods depends, indeed, on the quality of the separation between the lateral
$(\rho)$ and vertical $(z)$ components of the Coulomb interaction, with further
averaging over them.
We recall that the essence of the "removal of resonances" technique based
on the action-angle variables \cite{RRM}  is the identification of adiabatic
invariants, i.e., approximately conserved integrals of motion.
The averaging of the Hamiltonian over the fast variables is
a next step within this technique. In our case
such an invariant is  the action $J_z$, which is approximately conserved
due to large difference in a time scale between fast vertical ($z$) and
slow lateral ($xy$) dynamics. Therefore, the integration over fast
variable (angle) is well justified in this case.
As it was pointed out above, at small magnetic field a plain quantum-mechanical
procedure, which reduces to a variant of the first order perturbation theory,
is less justified.
In fact, the effective potential obtained from the plain quantum-mechanical procedure
is significantly stronger at small
distances between two electrons (see Appendix \ref{app1}). In turn, in this case
the effective coupling between the vertical and lateral confinements is stronger,
which indicates a less efficiency of the separation procedure.

The positions of the singlet-triplet transitions in
the same sample can be  reproduced as well with a vertical confinement
approximated by the square (Section \ref{sqw})  and
the triangular (Section \ref{tri}) potentials.
It appears from the behavior of the effective charge
for the considered models of the vertical confinement
(see Fig.~\ref{fig2}) that one obtains  equivalent results
if the ratio  $\kappa=b/(\Omega/\omega_z)$ between the distance
$b$ in the square/triangular well and the parabolic variable
$\Omega/\omega_z$ is $\kappa\sim 4$.
Note that this conclusion is independent on the particular
value of the vertical confinement in the parabolic model, since
this parameter enters in the calculations of the $k_\mathrm{eff}$
through the ratio $\Omega/\omega_z$.

Within the square well model the best agreement with the experimental data
is obtained for $a\approx 32.5$\,nm (see Fig.\ref{fig5}).
In virtue of the definition $b=\mu \Omega a^2/\hbar$ we have
\beq
\kappa=b/\biggl(\frac{\Omega}{\omega_z}\biggr) =
\frac{m^*a^2\hbar\omega_z}{2\hbar^2}\;.
\label{scal}
\eeq
Using the above value for $a$ and $\hbar\omega_z = 8$\,meV which
provides the equivalent result in the parabolic model, we obtain
$\kappa = 3.715$. Indeed, this estimation is close to the one extracted
from Fig.\ref{fig2} which is independent on the specific parameters of the
QD.

Similar agreement with the experimental data is obtained
by means of the triangular well potential ($b = \mu \Omega
z_m^2/\hbar$) for the electric field $F \approx 220$\,kV/m which
corresponds to $z_m \approx 32.1$\,nm (see Fig.~\ref{fig5}).
Note that in all considered models the effective charge
$k_\mathrm{eff}$ of the Coulomb interaction changes with the
increase of the magnetic field in a similar way (see
Fig.~\ref{fig5}) producing approximately the equivalent effect in
all three cases. At very large magnetic fields the effective charge
approaches almost the value $k_\mathrm{eff}\approx 0.95 k$.

Finally, a remark is in order.
For the sake of illustration we have used only
$k_\mathrm{eff}$ for the lowest basis states
($n_\rho = n_\rho^\prime = 0$) with different $m$.
However, even for the ground state
calculations with the aid of the exact diagonalization,
the interaction matrix elements
$ k_\mathrm{eff} \langle n_\rho^\prime,m|\rho^{-1}
|n_\rho^\prime,m\rangle$ with $n_\rho, n_\rho^\prime \ge 0$
have been taken into account. Namely, the diagonalization is
performed using the interaction matrix elements up to
$n_\rho, n_\rho^\prime = 10$.
Obviously, the accuracy of the method
would improve if one calculates
the 'effective charge matrix elements'
$k_{n_\rho,n_\rho^\prime}^{(m)} = \langle
n_\rho^\prime,m|\rho
V_\mathrm{int}^\mathrm{eff}|n_\rho^\prime,m\rangle$
for each interaction matrix element.
However, in this case the procedure would lose the simplicity and
become impractical. Fortunately,
from the comparison of the present results  with exact 3D
calculations we found that for the analysis of the ground state
properties it is sufficient to use only the elements $k_{0,0}^{(m)} \equiv
k_\mathrm{eff}^{(m)}$, even for the interaction matrix elements
with $n_\rho, n_\rho^\prime \ge 0$.

\section{Summary}
We developed the effective charge approach taking full account of the
thickness of two-electron quantum dots. Our approach is based on the
adiabatic approximation where the full
3D dynamics of two interacting electrons is separated by means of
action-angle variables on the independent vertical motion and the
lateral dynamics described by the effective 2D
Hamiltonian. The separation is reached due  to different time
scales in the vertical (fast) and lateral (slow) dynamics and
it is well justified in all types of QDs (vertical and lateral).
As a result, one must solve only the Schr\"odinger
equation for the 2D effective Hamiltonian where the full charge $k$ is replaced
by $k_\mathrm{eff}$ (see Eqs.(\ref{effch}), (\ref{intk})).
The eigenvalue problem was solved
by means of the exact diagonalization of the effective 2D Hamiltonian in
the Fock-Darwin basis.
To demonstrate the feasibility of the effective charge approach,
we considered
three different models for the vertical confinement: the parabolic,
the square  and triangular well potentials. The use of the
parabolic potential simplifies the analysis due to the
Kohn theorem which results in the analytical expression for
the effective charge, Eq.(\ref{kaap}).  For the square and
triangular well potentials we obtained  expressions for the
effective charge Eqs.(\ref{k2}),(\ref{ktr}),
respectively, that can be easily calculated numerically.

The value of the effective
charge depends on the (good)
quantum number $m$ of the correlated
state (see Fig.\ref{fig2}).
We established a scaling factor, Eq.(\ref{scal}), between the variables
used in the parabolic and the square/triangular well potentials.
Taking into account this factor, we found that the effective charge
(screening of the Coulomb interaction) affects quantum states for
all potentials in a similar way.
The screening due to the sample thickness is especially strong for quantum states
with small values of the quantum number $m$. We recall that
these states determine the structure of the ground state transitions
at small and intermediate values of the magnetic field.
Therefore, the screening provides a consistent way to
deal with the effect of the thickness upon the position of the
singlet-triplet transitions. In particular, the screening should be
taken into account
for the analysis of evolution of the energy difference between
singlet and triplet states in the magnetic field.
This energy is considered to be important
for analysis of the entanglement and concurrence in QDs (cf \cite{mar}).

The comparison of the results with available experimental data
\cite{ni} demonstrates a remarkable agreement and
lends support to the validity of the approach.
Being important for the states with small quantum number $m$,
the screening of the Coulomb interaction becomes small for the states with
large $m$, which dominate in the low-lying spectrum at large
magnetic fields. On the other hand, these states cause  strong
centrifugal forces which induce the electron
localization in a plane. In turn, the stronger is the magnetic field
the less important is the vertical confinement. It follows that  the 2D
bare Coulomb potential becomes reliable in 2D approaches
for the analysis of the ground state evolution of QDs
only at very large magnetic fields.

Thus, the 2D calculations with a bare Coulomb
interaction (with the charge $k$) overestimate the electron
correlations in two-electron QDs at small and intermediate values of the
magnetic field.
One may induce that similar conclusion could be valid
for QDs with more than two electrons. Notice, however, that
the confining frequency in the lateral plane
decreases with the increase of the electron number $N>2$ in
exact 3D calculations  at fixed vertical confinement
\cite{mel}, making QDs to be more of a two-dimensional system. A thorough analysis
of our approach for $N>2$ is left for the future.

\section*{Acknowledgments}
This work was partly supported
by Project No 141029 of Ministry of
Science and Environmental Protection of Serbia,
by Grant No. FIS2005-02796 (MEC, Spain) and
RFBR Grant No. 08-02-00118 (Russia).

\appendix

\section{Effective interaction potential
for the parabolic vertical confinement at $n_z=0$}
\label{app1}

\begin{figure}
\includegraphics[scale=.5]{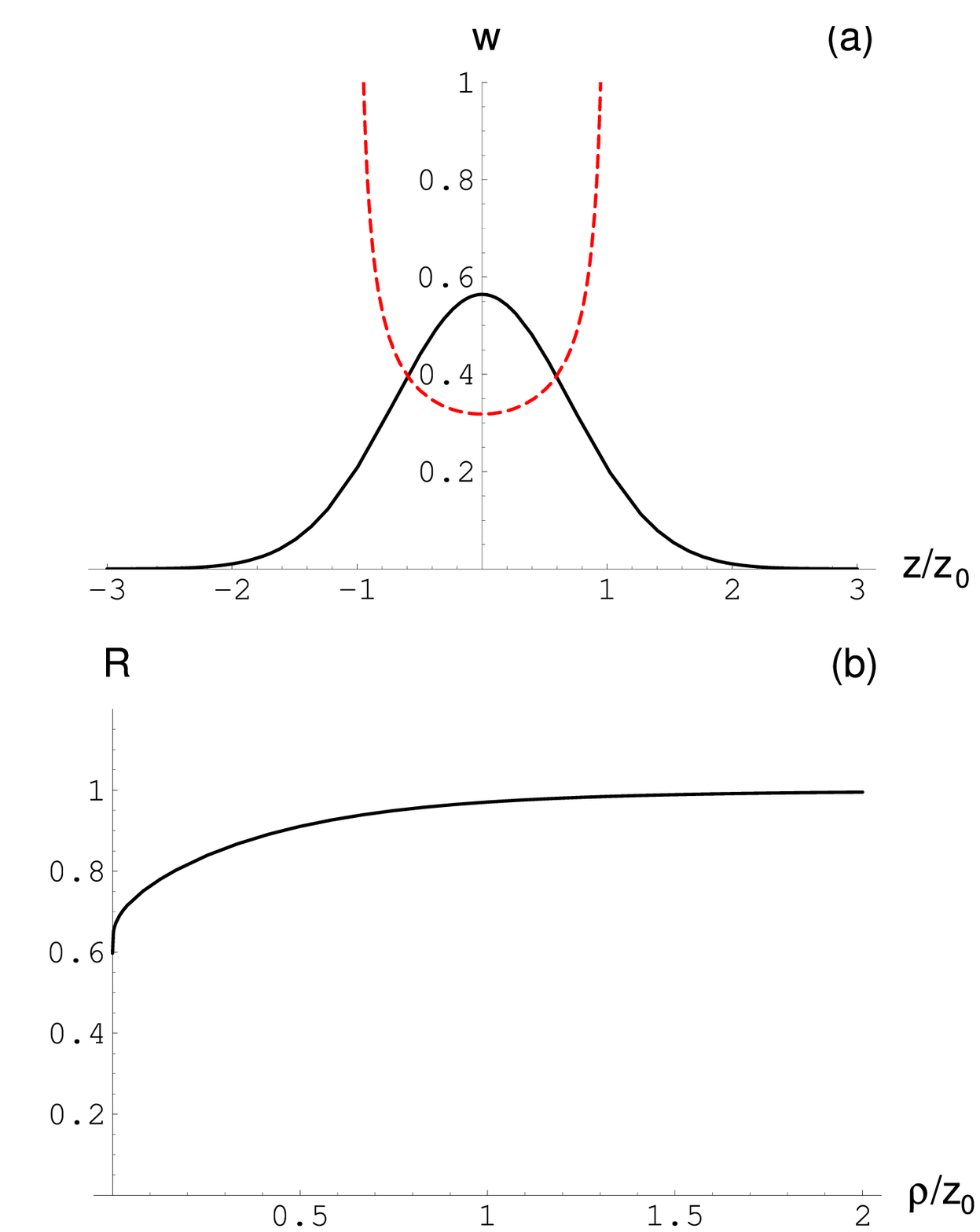}
\caption{(Color online)(a) The weight functions $w_\mathrm{qm}$ (solid
line), $w_\mathrm{ad}$ (dashed line) and (b) the ratio $R =
V_\mathrm{int}^\mathrm{eff(ad)}/V_\mathrm{int}^\mathrm{eff(qm)}$ for the parabolic vertical
confinement.}
\label{fig6}
\end{figure}

The effective electron-electron interaction potential is obtained by averaging the
Coulomb term $V_C(\rho,z)$ in the z-direction
\beq
V_\mathrm{int}^\mathrm{eff}(\rho) = \int w(z) V_C(\rho,z)\, dz,
\eeq
where $w(z)$ is the corresponding weight function.
By dint of the quantum-mechanical averaging one obtains the effective potential
\beq
V_\mathrm{int}^\mathrm{eff(qm)}(\rho) = \int_{-\infty}^{\infty}
|\psi_{n_z}(z)|^2 V_C(\rho,z)\, dz,
\label{vqm}
\eeq
while in the adiabatic approach we have the effective potential
\begin{eqnarray}
V_\mathrm{int}^\mathrm{eff(ad)}(\rho) &=& \frac{1}{\pi}\int_{-\pi/2}^{\pi/2}
V_C(\rho,z(\theta_z))\, d\theta_z\nonumber \\
&=& \frac{1}{\pi} \int_{-z_0}^{z_0}
\bigg(\frac{d\theta_z}{dz}\bigg) V_C(\rho,z)\, dz.
\label{vcl}
\end{eqnarray}
For $n_z = 0$,  we have for the quantum-mechanical weight function
\beq
w_\mathrm{qm} \equiv |\psi_{n_z=0}(z)|^2=
\frac{\mathrm{e}^{-(z/z_0)^2}}{\sqrt{\pi}z_0},
\label{wqm}
\eeq
while, by virtue of Eq.(\ref{ztz}) and $J_z=\hbar/2$, one obtains
for the adiabatic weight function
\beq w_\mathrm{ad}\equiv \pi^{-1} (d\theta_z/dz) =
\frac{1}{\pi z_0\sqrt{1-(z/z_0)^2}},
\label{wcl}
\eeq
where $z_0 = \sqrt{\hbar/\mu\omega_z}$.
 As it can be seen from
Fig.~\ref{fig6}(a) the weight functions $w_\mathrm{qm}$ (solid line) and
$w_\mathrm{ad}$ (dashed line) are completely
different and, therefore, one may expect different contributions of the
z-motion in the effective 2D potential, i.e., a different
$\rho$-$z$ coupling.

By dint of the definition $V_C = k/(\rho^2+z^2)^{1/2}$ and
Eqs.(\ref{wqm}),(\ref{wcl}) one obtains
for the effective interaction potentials (Eqs.(\ref{vqm}),(\ref{vcl})):
\bea
V_\mathrm{int}^\mathrm{eff(qm)}(\rho) & = & \frac{k}{\sqrt{\pi}z_0}\,
\mathrm{e}^{\rho^2/2z_0^2} K_0\biggl(\frac{\rho^2}{2z_0^2}\biggr),\\
%
V_\mathrm{int}^\mathrm{eff(ad)}(\rho) & = &
\frac{2k}{\pi\rho}\,K\biggl(-\frac{z_0^2}{\rho^2}\biggr),
\eea
%
where $K_0$ and $K$ are the the modified Bessel function of the
2nd kind and the complete elliptic integral of the first kind,
respectively.

Although the potential $V_C^\mathrm{eff(qm)}$  is akin to the potential
$V_\mathrm{int}^\mathrm{eff(ad)}$, the ratio $R =
V_\mathrm{int}^\mathrm{eff(ad)}/V_\mathrm{int}^\mathrm{eff(qm)}< 1$
 at $0<\rho<\sqrt{2}z_0$ (see Fig.~\ref{fig6}(b)).
We recall that for the  lowest limit of the sample thickness
$a = 2z_m =\sqrt{2}z_0$ (see Fig.\ref{fig1}).
In other words, when electrons are close to each other (for $\rho < a$),
the effective quantum-mechanical potential
is significantly stronger than the adiabatic one.
At distances between electrons $\rho > a$ two potentials  reach each
other asymptotically ($R \to 1$).

\section{Action-angle variables in the one-dimensional
square well potential}
\label{appA}

If $\varepsilon$ is the energy of a particle of the mass $m^*$ moving in
the infinite square well potential (\ref{sqwell}), its momentum
$p_z = \pm\sqrt{2m^*\varepsilon}$ change the sign at the turning points
$\pm a/2$. In this case the action variable is
\beq
J_z = \frac{1}{\pi}\int_{-a/2}^{a/2} p_z\,dz =
\frac{a}{\pi}\sqrt{2m^* \varepsilon} \;.
\eeq
Accordingly, the energy expressed in terms of $J_z$ is
$\varepsilon = \pi^2 J_z^2/(2m^*a^2)$.
By dint of the semiclassical quantization condition $J_z =
\hbar\,(n_z + 1)$ ($n_z = 0,1,2,...$), one obtains
energy levels for a particle moving in the potential
(\ref{sqwell}). The value $1$ in this condition is
a Maslov index (cf \cite{brack}).
It is convenient to define $n_z \to n_z+1$.
Finally, one has
\beq
\varepsilon = \frac{\hbar^2}{2m^*}\biggl(\frac{\pi n_z}{a}\biggr)^2,
\quad n_z=1,2,...
\label{ebox}
\eeq

Integrating equation $m^*{\dot z} = p_z$,
from the condition $z(T/4) = a/2$ (for $-a/2 < z < a/2$)
one can define the period $T = 2a\sqrt{m^*/2\varepsilon}$.
It follows that the angle variable $\theta_z$ has the form
\beq
\theta_z =\omega_z t = \frac{2\pi}{T}t =
\frac{\pi}{a}\sqrt{\frac{2\varepsilon}{m^*}}\,t =\frac{\pi}{a}\,z\;.
\label{awell}
\eeq

\section{Action-angle variables in a
triangular well potential}
\label{appB}

For a particle moving with energy $\varepsilon$
in the triangular well potential (\ref{triwell})
the turning points are $z = 0$ and $z = z_m$.
In virtue of equation
$p_z \equiv \pm \sqrt{2m^*(\varepsilon - eFz)} = 0$
one defines the variable $z_m = \varepsilon/eF$ and, respectively,
the action variable
\beq
J_z = \frac{1}{\pi}\int_0^{z_m} p_z\,dz =
\frac{2\sqrt{2m^*} \varepsilon^{3/2}}{3\pi eF}\;.
\label{triact}
\eeq
In turn, the energy expressed in terms of $J_z$ is
\beq
\varepsilon=[3\pi J_zeF/\sqrt{8m^*}]^{2/3}\;.
\label{tren}
\eeq

From Eq.(\ref{tren}) one can define the angle variable
\beq
\theta_z = \omega_z t=\frac{\partial \varepsilon}{\partial J_z}t=
\frac{\pi eF}{\sqrt{2m^*\varepsilon}}\,t
\label{atr}
\eeq
and, consequently, the period
$T = 2\pi/\omega_z = 2\sqrt{2m^* \varepsilon}/eF$.
Integrating equation $m^*{\dot z} = p_z$, one obtains
\beq t = m^*\int_0^z \frac{dz}{p_z} = \frac{\sqrt{2m^* \varepsilon}}{eF}
\biggl(1-\sqrt{1-\frac{eF}{\varepsilon}z}\biggr), \label{t-z} \eeq
for $0 \le z\le z_m$ or
\beq
z = \sqrt{\frac{2\varepsilon}{m^*}}\,t - \frac{eF}{2m^*}\,t^2, \quad
0 \le t \le T/2 .
\eeq
By dint of equation $t =\theta_z/\omega_z$, it is useful to
establish the relation between the coordinate $z$ and the
variable $\theta_z$
\beq
z = z_m \biggl[ 2\frac{\theta_z}{\pi} -
\biggl(\frac{\theta_z}{\pi}\biggr)^2\biggr], \quad 0 \le \theta_z
\le \pi.
\label{zang}
\eeq
For the triangular potential one has
the semiclassical quantization condition $J_z =
\hbar\,(n_z + 3/4)$, where $n_z = 0,1,2,...$ and the Maslov index $=3/4$.
Redefining $n_z\to n_z-1$, one obtains the semiclassical energy levels
from Eq.(\ref{tren})
\beq
\varepsilon_{n_z}=c_{n_z}\biggl[\frac{(eF\hbar)^2}{2m^*}\biggr]^{1/3}, \quad
c_{n_z}=[\hbox{$\frac{3}{2}$}\pi
(n_z-\hbox{$\frac{1}{4}$})]^{2/3}
\label{etr}
\eeq
with $n_z=1,2,...$.

\end{document}